\documentclass[conference]{IEEEtran}
\usepackage{epsfig}
\usepackage{amssymb}
\usepackage{latexsym,graphicx}
\usepackage{longtable}
\usepackage{algpseudocode} 
\usepackage{algorithm}
\usepackage{array}
\usepackage{mathrsfs}
\usepackage{color}
\usepackage{amsmath,graphicx}
\usepackage{epsfig}
\usepackage{amssymb}
\usepackage{setspace}
\usepackage{cite}
\usepackage{hyperref}
\usepackage{amssymb}
\usepackage{caption}
\usepackage{subcaption}
\def\BibTeX{{\rm B\kern-.05em{\sc i\kern-.025em b}\kern-.08em
    T\kern-.1667em\lower.7ex\hbox{E}\kern-.125emX}}
 \makeatletter
 \let\NAT@parse\undefined
 \makeatother

\begin{document}
%
% paper title
\title{Multi-Antenna Assisted Virtual Full-Duplex Relaying with Reliability-Aware Iterative Decoding} 
%
%
% author names and IEEE memberships
% note positions of commas and nonbreaking spaces ( ~ ) LaTeX will not break
% a structure at a ~ so this keeps an author's name from being broken across
% two lines.
% use \thanks{} to gain access to the first footnote area
% a separate \thanks must be used for each paragraph as LaTeX2e's \thanks
% was not built to handle multiple paragraphs
\author{$\text{Jiancao Hou}^{\dagger}$, $\text{Sandeep Narayanan}^{\dagger}$, $\text{Yi Ma}^{\star}$, and $\text{Mohammad Shikh-Bahaei}^{\dagger}$
% <-this % stops a space
\\{\small $^{\dagger}$Centre for Telecommunications Research, King's College London, London, UK}
\\{\small $^{\star}$Institute for Communication Systems, University of Surrey, Guildford, UK}
\\{\small e-mail:\{jiancao.hou, sandeep.kadanveedu, m.sbahaei\}@kcl.ac.uk, y.ma@surrey.ac.uk}% <-this % stops a space
%\thanks{This work was partly funded by the ICT-248894 WHERE 2.}
% <-this % stops a space
%\thanks{This work was supported in part by the Engineering and Physical Science Research Council (EPSRC) through the SENSE grant EP/P003486/1.

%J. Hou, S. Narayanan and M. Shikh-Bahaei are with the Department of Informatics, King's College London, London, United Kingdom, WC2R 2LS. E-mail: \{jiancao.hou, sandeep.kadanveedu, m.sbahaei\}@kcl.ac.uk.

%N. Yi and Y. Ma are with the Institute for Communication Systems, University of Surrey, Guildford, United Kingdom, GU2 7XH. E-mail: \{n.yi, y.ma\}@surrey.ac.uk.}
}
\markboth{} {Shell \MakeLowercase{\textit{et al.}}: Bare Demo of
IEEEtran.cls for Journals}\maketitle

\begin{abstract}
%Summary of your paper: brief introduction, motivation, and contributions.
In this paper, a multi-antenna assisted virtual full-duplex (FD) relaying with reliability-aware iterative decoding at destination node is proposed to improve system spectral efficiency and reliability. This scheme enables two half-duplex relay nodes, mimicked as FD relaying, to alternatively serve as transmitter and receiver to relay their decoded data signals regardless the decoding errors, meanwhile, cancel the inter-relay interference with QR-decomposition. Then, by deploying the reliability-aware iterative detection/decoding process, destination node can efficiently mitigate inter-frame interference and error propagation effect at the same time. Simulation results show that, without extra cost of time delay and signalling overhead, our proposed scheme outperforms the conventional selective decode-and-forward (S-DF) relaying schemes, such as cyclic redundancy check based S-DF relaying and threshold based S-DF relaying, by up to 8 dB in terms of bit-error-rate.
\end{abstract}

%\begin{keywords}
%Iterative receiver, maximum a posteriori (MAP), successive erroneous relaying, single-input multiple-output (SIMO).
%\end{keywords} 

\IEEEpeerreviewmaketitle
%%%%%%%%%%%%%%%%%%%%%%%%%%%%%%%%%%%%%%%%%%%%%%%%%%%%%%%%%%%%%%%%%%%%%%
 
                         %%%I. Introduction%%% 

%%%%%%%%%%%%%%%%%%%%%%%%%%%%%%%%%%%%%%%%%%%%%%%%%%%%%%%%%%%%%%%%%%%%%%
\section{Introduction}
%Para. 1 introduces the background related to your research.
Next generation wireless communication networks are expected to rely on low latency and high spectral efficiency cooperative relaying nodes \cite{Orikumhi2017}. This is due to their capability to serve as a virtual multi-antenna system to combat fading and be able to extend communication coverage\cite{Laneman2004,Azarian2005,Shadmand2010}. However, conventional cooperative relaying networks limit relay nodes on half-duplex (HD) mode, where the HD relay nodes can either receive or transmit data symbols at a given time-instant. With such kind of design, the cooperative system can avoid inter-frame interference and exploit spatial diversity gain. On the other hand, it suffers from spectral efficiency loss since transmission of one data frame needs to occupy two independent time slots or frequency bands. 

Recent research works in \cite{Sabharwal2014,Liu2015,Naslcheraghi2017} show that full-duplex (FD) relaying becomes feasible for simultaneous transmission and reception when advanced self-interference cancellation methods can be exploited \cite{Hong2014,Ahmed2015}. However, in practical environments, self-interference cannot be completely cancelled due to inaccurate self-interference channel modeling and/or limited dynamic range of analog-to-digital converter\cite{Kim2015}. Consequently, the residual self-interference increases burden on decoding process at the FD relay node and results in error propagation. In \cite{Oechtering2004,Fan2007,Lehmann2016}, a promising technique named virtual FD relaying (or two-path successive relaying) has been proposed, where two HD relay nodes mimic a FD relay node and alternately serve as transmitter and receiver to continuously transmit data frames for every channel use. This scheme benefits from simplified curcit designs, and the optimal performance can be achieved if both HD relay nodes are able to perfectly cancel the inter-relay interference and decode their received messages\cite{Wicaksana2011,Fan2017}. Otherwise, error propagation effects degrade system performances. 

In order to avoid error propagation effects, the authors in \cite{Fan2007,Wicaksana2011,Bobarshad2009,Zhang2015,Olivo2016} presented adaptive retransmission based relaying schemes to guarantee perfect decoding at relay nodes. However, this scheme has additional costs of time delay and signalling overhead \cite{Shikh-Bahaei2004,Olfat2005,Kobravi2007,Shadmand2009}. Alternatively, the authors in \cite{Al-Habian2011,Kwon2010,Yi2015} proposed symbol-level selective transmission schemes, where relay nodes can predict their correctly decoded symbols based on the pre-determined threshold and discard the the erroneously decoded symbols, resulting in fewer errors being forwarded to destination node. However, in order to guarantee the quality of decoding process at destination, the relay nodes need to forward the position of their discarded symbols to destination node, where the additional signalling overhead still exists. On the other hand, imperfect inter-relay interference cancellation may reduce the decoding capability at relay nodes \cite{Sun2011,Basar2014,Zarringhalam2009}. With the help of multiple receive-antenna, the detection techniques, such as zero-forcing (ZF) or minimum mean square error (MMSE), can be implemented to suppress the inter-relay interference. However, the majority drawback of ZF is noise power boost, and MMSE needs to know noise variance information.

Motivated by the above discussion, in this paper, we propose a multi-antenna assisted virtual FD relaying with reliability-aware iterative decoding at destination node in order to improve system spectral efficiency and reliability. Specifically, relay nodes first ultilize QR decodposition method to perfectly cancel the inter-relay interference and then forward their decoded data frames regardless the decoding errors. Subsequently, destination node performs the reliability-aware iterative detection/decoding method by taking the estimated probability of decoding error at relay nodes (e.g. $p_e$) into account. In the literature, multi-antenna assisted iterative decoding was proposed in \cite{Vikalo2004} for point-to-point communication network and in \cite{Lee2009} for network coding aided two-way relaying network. Comparison with the works in \cite{Vikalo2004,Lee2009}, our proposed scheme is for virtual FD relaying network and the additional challenge is how to jointly cancel inter-frame interference and enjoy spatial diversity gain at destination node. In addition, we also introduce an estimation method of $p_e$ for the iterative decoding process. Simulation results show that, without extra cost of time delay and signalling overhead , our proposed scheme exhibits up to 8 dB bit-error-rate (BER) performance gain by comparing with conventional cyclic redundancy check (CRC) based S-DF relaying and threshold based S-DF relaying schemes.

% or exploiting the source-relay (S-R) link correlation information at destination \cite{Anwar2012}. The first strategy is implemented if relays can be intricately designed. However, in certain cases, the relays should be designed with low complexity, thus the second strategy is the better choice. This letter will focus on the second strategy. In \cite{Anwar2012}, the probability of decoding error at relays is deemed as the S-R link correlation information, and such information can be embedded into the iterative decoding at destination to mitigate the error propagation effects.  

%%%%%%%%%%%%%%%%%%%%%%%%%%%%%%%%%%%%%%%%%%%%%%%%%%%%%%%%%%%%%%%%%%%%%%

                     %%%II. System Model%%% 

%%%%%%%%%%%%%%%%%%%%%%%%%%%%%%%%%%%%%%%%%%%%%%%%%%%%%%%%%%%%%%%%%%%%%%
\section{System Model}
\begin{figure}[t] 
\begin{center}
\epsfig{figure=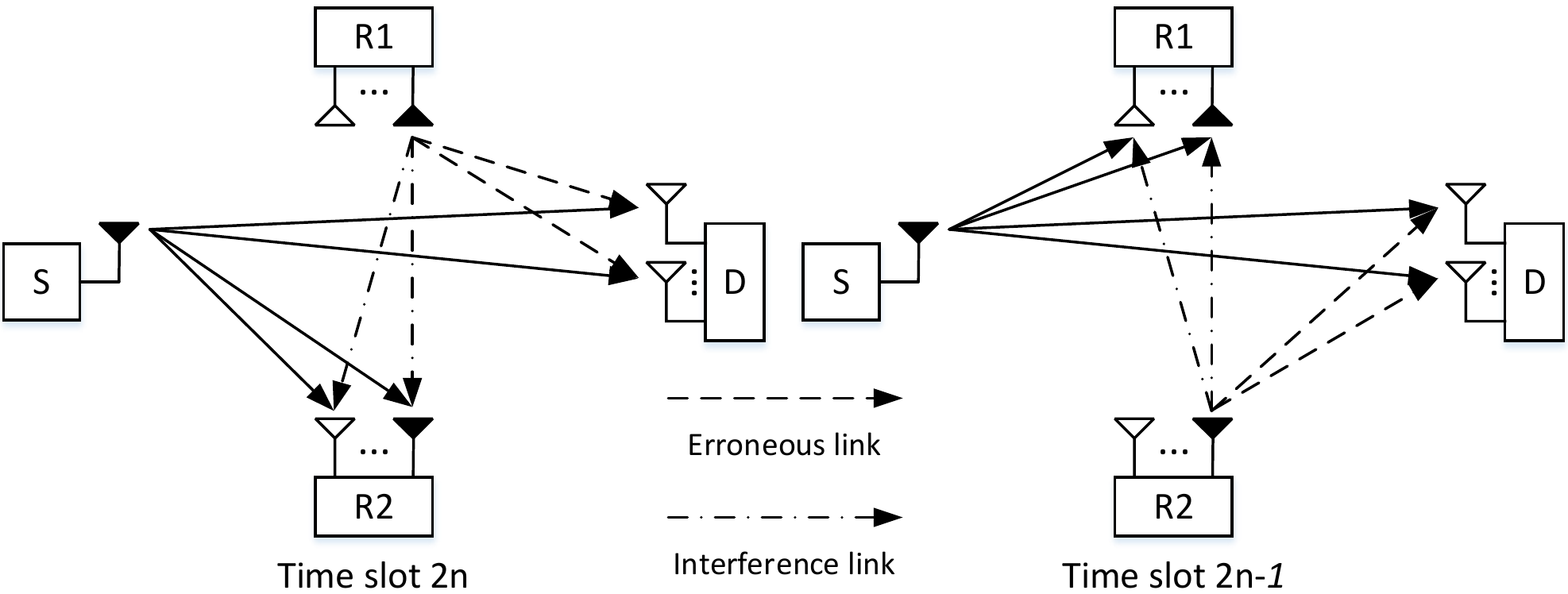,scale=0.43,angle=0}
\end{center}
\caption{Illustration of the virtrual FD relaying network.}\label{F1}
\end{figure}

% Provide a comprehensive description of the system model which leads to your theoretical analysis. A block diagram is usually demanded to clarify your % presentation. Formulate the problem you want to resolve. Please note this system model should be much related to your simulations.
Fig.~\ref{F1} illustrates the multi-antenna assisted virtual FD relaying network with one single-antenna based source node (S), two $N$-antenna based HD relay nodes (R1/R2) and one $N$-antenna based destination node (D). 
All transmissions are using single antenna and all receptions are using $N$ antennas.\footnote{In order to keep consistent with the source node, two relay nodes transmit data frames with single antenna. The multi-antenna based transmission at relay nodes can be easily extended to enjoy additional antenna gain.} We also assume discrete-time block fading channels, which remain static over each time slot. All reception nodes can estimate and have their related channel state information. The source node first encodes the information bits $\mathbf{b}$ using the turbo-like encoders, e.g., $m$-rate serial concatenated convolutional codes, to generate the coded bits $\mathbf{c}$. Then, $\mathbf{c}$ are mapped into $\mathbf{x}$ transmission symbols based on $Q$-ary modulation scheme. Subsequently, $\mathbf{x}$ are divided into $L$ frames, and without loss of generality $L$ is assumed to be even. 

As shown in Fig.~\ref{F1}, the source node continuously transmits the $l^{\mathrm{th}}$ frame in the $l^{\mathrm{th}}$ time slot, where $l=1,2,\ldots,L$. In the first time slot, only relay node one (R1) and destination node receive the first data frame from source node. Starting from the second time slot, two relay nodes alternatively transmit and receive data frames until the $L^{\mathrm{th}}$ time slot. Then, in the $(L+1)^{\mathrm{th}}$ time slot, only relay node two (R2) forwards its last decoded frame (i.e. the $L^{\mathrm{th}}$) to destination node. Let $\mathbf{Y}_{\mathrm{R}j}(l)\in\mathcal{C}^{N\times M}$ and $\mathbf{Y}_{\mathrm{D}}(l)\in\mathcal{C}^{N\times M}$ be the received signal matrices at the $j^{\mathrm{th}},j\in\{1,2\}$, relay and the destination node in the $l^{\mathrm{th}}$ time slot, respectively, where $M$ is the number of symbols per frame. Then, we have
\begin{equation}\label{eq01}
\mathbf{Y}_{\mathrm{R}j}(l)=\mathbf{H}_{\mathrm{R}j}(l)\mathbf{X}(l)+\mathbf{V}_{\mathrm{R}j}(l), 
\end{equation}
\begin{equation}\label{eq02}
\mathbf{Y}_{\mathrm{D}}(l)=\mathbf{H}_{\mathrm{D}}(l)\mathbf{X}(l)+\mathbf{V}_{\mathrm{D}}(l),
\end{equation}
where $\mathbf{X}(l)\triangleq[\mathbf{x}_{\mathrm{R}\overline{j}}(l-1),\mathbf{x}_{\mathrm{S}}(l)]^{T}$ is the transmitted signal matrix in the $l^{\mathrm{th}}$ time slot composed by the $(l-1)^{\mathrm{th}}$ frame sent from the $\overline{j}^{\mathrm{th}},\overline{j}\neq j\in\{1,2\}$, relay, i.e., $\mathbf{x}_{\mathrm{R}}(l-1)\in\mathcal{C}^{M}$, and the $l^{\mathrm{th}}$ frame sent from the source node, i.e., $\mathbf{x}_{\mathrm{S}}(l)\in\mathcal{C}^{M}$; $(\cdot)^{T}$ is transpose of a matrix; $\mathbf{H}_{\mathrm{R}j}(l)\in\mathcal{C}^{N\times2}$ is the channel matrix which consists of the R-R link and S-R link in the $l^{\mathrm{th}}$ time slot; $\mathbf{H}_{\mathrm{D}}(l)\in\mathcal{C}^{N\times2}$ is the channel matrix which consists of R-D link and S-D link in the $l^{\mathrm{th}}$ time slot; $\mathbf{V}_{\mathrm{R}j}(l)\in\mathcal{C}^{N\times M}$ and $\mathbf{V}_{\mathrm{D}}(l)\in\mathcal{C}^{N\times M}$ are the additive white Gaussian noise (AWGN) matrices at the $j^{\mathrm{th}}$ relay and the destination nodes in the $l^{\mathrm{th}}$ time slot, respectively. It is worth noting that, in the first time slot, since relay node one receives data frame sent from the source node without inter-relay interference effect, the transmitted signal in this case should be $\mathbf{X}(1)=[\mathbf{0},\mathbf{x}_{\mathrm{S}}(1)]^{T}$, where $\mathbf{0}$ is a column vector with all zeros. Similarly, in the $(L+1)^{\mathrm{th}}$ time slot, since only relay node two (R2) sends data frame to destination node, the transmitted signal in this case should be $\mathbf{X}(L+1)=[\mathbf{x}_{\mathrm{R}2}(L),\mathbf{0}]^{T}$. Apart from that, in this paper, relay nodes decodes its received frame and then forwards its re-encoded version regardless of decoding errors, then the destination node will based on our proposed method to mitigate the error propagation effect. 

%%%%%%%%%%%%%%%%%%%%%%%%%%%%%%%%%%%%%%%%%%%%%%%%%%%%%%%%%%%%%%%%%%%%%%

                     %%%III. Iterative Sub-optimal Power Allocation Approach Design%%% 

%%%%%%%%%%%%%%%%%%%%%%%%%%%%%%%%%%%%%%%%%%%%%%%%%%%%%%%%%%%%%%%%%%%%%%
\section{The Proposed Scheme for Interference and Error Propagation Mitigation}
According to the system model described above, there are three main factors that affect the system performances, which are the inter-relay interference generated at relay nodes, the inter-frame interference generated at destination node, and the error propagation effect due to decoding errors at relay nodes. In this section, we mainly introduce how to cancel and limit these impacts with our proposed methods.

%%%%%%%%%%%%%%%%%%%%%%%%%%%%%%%%%%%%%%%%%%%%%%%%%%%%%%%%%%%%%%%%%%%%%%
\subsection{Inter-relay Interference Cancellation at Relay Nodes}
Due to alternative transmission and reception between two relay nodes, the presented inter-relay interference should be cancelled before they can decode the received data frames. In this case, by introducing QR decomposition, the inter-relay interference at relay nodes can be perfectly cancelled without boosting noise power. Specifically, we take the $j^{\mathrm{th}}$ relay node as an example and assume $\mathbf{Q}_{\mathrm{R}j}(l)\in\mathcal{C}^{N\times N}$ is the unitary matrix and $\mathbf{R}_{\mathrm{R}j}(l)\in\mathcal{C}^{N\times2}$ is the upper triangular matrix, where both matrices are generated from QR decomposition of the channel matrix $\mathbf{H}_{\mathrm{R}j}(l)$. Then, the received signal at the $j^{\mathrm{th}}$ relay node, after multiplying the unitary matrix $\mathbf{Q}_{\mathrm{R}j}(l)$, can be expressed as
\begin{eqnarray}\label{eq03}
\hat{\mathbf{Y}}_{\mathrm{R}j}(l)&=&\mathbf{Q}^{H}_{\mathrm{R}j}(l)\mathbf{Y}_{\mathrm{R}j}(l), \nonumber\\
&=&\mathbf{R}_{\mathrm{R}j}(l)\left[\begin{array}{cc}\mathbf{x}^{T}_{\mathrm{R}\overline{j}}(l-1) \\
\mathbf{x}^{T}_{\mathrm{S}}(l)
\end{array}\right]+\mathbf{Q}^{H}_{\mathrm{R}j}(l)\mathbf{V}_{\mathrm{R}j}(l),
\end{eqnarray}
where $(\cdot)^{H}$ is hermitian of a matrix. Due to the upper triangular structure of $\mathbf{R}_{\mathrm{R}j}(l)$, the desired signal sent from the source node without inter-relay interference effects is the second row of matrix $\hat{\mathbf{Y}}_{\mathrm{R}j}(l)$. In addition, due to the nice feature of $\mathbf{Q}_{\mathrm{R}j}(l)$, the multiplication of $\mathbf{V}_{\mathrm{R}j}(l)$ by $\mathbf{Q}_{\mathrm{R}j}(l)$ does not change the power of AWGN matrix. Thus, based on the above description, the $j^{\mathrm{th}}$ relay node can decode the second row of $\hat{\mathbf{Y}}_{\mathrm{R}j}(l)$ in order to regenerate data frame sent from source node, i.e., $\mathbf{x}_{\mathrm{R}j}(l)$. Then, it will be forwarded to destination node in the $(l+1)^{\mathrm{th}}$ time slot. 

%%%%%%%%%%%%%%%%%%%%%%%%%%%%%%%%%%%%%%%%%%%%%%%%%%%%%%%%%%%%%%%%%%%%%%
\subsection{Reliability-Aware Iterative Decoding at Destination Node}
In this paper, we allow relay nodes forward their received signals regardless decoding error. In this case, error propagation could degrade the system performance. In addition, due to simultaneously receiving signals sent from source and one of relay nodes, the inter-frame interference at destination node also affects system performance. In this subsection, we introduce a reliability-aware iterative detection/decoding method to mitigate the above effects. 

It is shown in the literature that MAP based detection technique has been widely used to cancel inter-frame interference. However, in order to allow the MAP detector have the capacity to jointly cancel inter-frame interference and remove the error propagation effects, we need to modify its detection process. Specifically, we start from transforming \eqref{eq02} to the real-valued equivalent form. Such transformation allows the proposed method suit for different modulation schemes, e.g., generalized $Q$-ary modulation schemes. Let's assume that $\tilde{\mathbf{Y}}_{\mathrm{D}}(l)\in\mathcal{R}^{2N\times M}$, $\tilde{\mathbf{X}}(l)\in\mathcal{R}^{2N\times M}$, and $\tilde{\mathbf{V}}_{\mathrm{D}}(l)\in\mathcal{R}^{2N\times M}$ are the real matrices transformed from $\mathbf{Y}_{\mathrm{D}}(l)$, $\mathbf{X}(l)$, and $\mathbf{V}_{\mathrm{D}}(l)$, respectively, where we have
\begin{equation}\label{eq04}
\tilde{\mathbf{Y}}_{\mathrm{D}}(l)=[\mathcal{R}(\mathbf{Y}_{\mathrm{D}}(l))^{T},\mathcal{I}(\mathbf{Y}_{\mathrm{D}}(l))^{T}]^{T}, 
\end{equation}
\begin{equation}\label{eq05}
\tilde{\mathbf{X}}(l)=[\mathcal{R}(\mathbf{X}(l))^{T},\mathcal{I}(\mathbf{X}(l))^{T}]^{T}, 
\end{equation}
\begin{equation}\label{eq06}
\tilde{\mathbf{V}}_{\mathrm{D}}(l)=[\mathcal{R}(\mathbf{V}_{\mathrm{D}}(l))^{T},\mathcal{I}(\mathbf{V}_{\mathrm{D}}(l))^{T}]^{T}. 
\end{equation}
Additionally, let $\tilde{\mathbf{H}}_{\mathrm{D}}(l)\in\mathcal{R}^{2N\times4}$ denote the real-valued channel matrix transformed from $\mathbf{H}_{\mathrm{D}}(l)$, as
\begin{eqnarray}\label{eq07}
\tilde{\mathbf{H}}_{\mathrm{D}}(l)=\left[\begin{array}{cc}
\mathcal{R}(\mathbf{H}_{\mathrm{D}}(l)) & -\mathcal{I}(\mathbf{H}_{\mathrm{D}}(l)) \\
\mathcal{I}(\mathbf{H}_{\mathrm{D}}(l)) & \mathcal{R}(\mathbf{H}_{\mathrm{D}}(l)) \end{array} \right]. 
\end{eqnarray}
Then, the real-valued equivalent form of \eqref{eq02} can be expressed as
\begin{equation}\label{eq08}
\tilde{\mathbf{Y}}_{\mathrm{D}}(l)=\tilde{\mathbf{H}}_{\mathrm{D}}(l)\tilde{\mathbf{X}}(l)+\tilde{\mathbf{V}}_{\mathrm{D}}(l). 
\end{equation}

According to the work in \cite{Vikalo2004}, the conventional MAP detector is to solve the optimization problem, which is
\begin{equation}\label{eq09}
\min_{\tilde{\mathbf{x}}^{(m)}(l)}\Bigg[\underbrace{\|\tilde{\mathbf{y}}^{(m)}_{\mathrm{D}}(l)-\tilde{\mathbf{H}}_{\mathrm{D}}(l)\tilde{\mathbf{x}}^{(m)}(l)\|^2}_{\triangleq\Delta^{(m)}(l)}-\sum^{2N}_{k=1}{\mathrm{log}p(\tilde{x}^{(l)}_{k,m})}\Bigg],
\end{equation}
where $\tilde{\mathbf{x}}^{(m)}(l)$ and $\tilde{\mathbf{y}}^{(m)}_{\mathrm{D}}(l)$ are the $m^{\mathrm{th}}$ columns of $\tilde{\mathbf{X}}(l)$ and $\tilde{\mathbf{Y}}_{\mathrm{D}}(l)$, respectively;\footnote{It is worth noting that $\tilde{\mathbf{x}}^{(m)}(l)$ in \eqref{eq09} are the corresponding trial bits vector used in this hypothesis-detection problem.} $p(\tilde{x}^{(l)}_{k,m})$ is the \textit{a posteriori} probability for $\tilde{x}^{(l)}_{k,m}$, where $\tilde{x}^{(l)}_{k,m}$ is the $k^{\mathrm{th}}$ component of the symbol vector $\tilde{\mathbf{x}}^{(m)}(l)$. As each constellation symbol represents $\log_{2}Q$ modulated bits (e.g., for a $Q$-ary constellation), we assume $c^{(l)}_{i,m},\forall i\in[1,\ldots,2\log_{2}Q],$ are the modulated bits for the symbol vector $\tilde{\mathbf{x}}^{(m)}(l)$.\footnote{Two times of $\log_{2}Q$ represents two symbols included in $\tilde{\mathbf{x}}^{(m)}(l)$, where one is from the source node, and one is from the relay node.} Since the output of the MAP detector requires the soft information for the next step decoding process, the LLR of the $i^{\mathrm{th}}$ modulated bit for the symbol vector $\tilde{\mathbf{x}}^{(m)}(l)$ can be expressed as
\begin{eqnarray}\label{eq10}
L[c^{(l)}_{i,m}|\tilde{\mathbf{y}}^{(m)}_{\mathrm{D}}(l)]=\log\frac{\mathrm{Pr}[c^{(l)}_{i,m}=+1|\tilde{\mathbf{y}}^{(m)}_{\mathrm{D}}(l)]}{\mathrm{Pr}[c^{(l)}_{i,m}=-1|\tilde{\mathbf{y}}^{(m)}_{\mathrm{D}}(l)]}~~~~~~~~~~~~~\nonumber\\
=\mathrm{log}\frac{\sum_{\mathbf{x}:c^{(l)}_{i,m}=+1}e^{-\Delta^{(m)}(l)+\sum_{j}{\mathrm{log}p(c^{(l)}_{j,m})}}}{\sum_{\mathbf{x}:c^{(l)}_{i,m}=-1}e^{-\Delta^{(m)}(l)+\sum_{j}{\mathrm{log}p(c^{(l)}_{j,m})}}},
\end{eqnarray}
where $\Delta^{(m)}(l)$ has been defined in \eqref{eq09}; $\mathbf{x}$ is the short-hand used to denote the symbol vector $\tilde{\mathbf{x}}^{(m)}(l)$; $c^{(l)}_{i,m}=+1$ represent logical 0 with amplitude level +1, and $c^{(l)}_{i,m}=-1$ represent logical 1 with amplitude level -1.

Different from the conventional MAP detector presented above, our proposed MAP detector needs to be aware of the error propagation effects. Specifically, we first define $p^{(l)}_e$ as the probability of decoding error at one of relay nodes in the $l^{\mathrm{th}}$ time slot. This probability can be estimated during the iterative decoding at destination node. More detailed analysis of its estimation method will be introduced in Section III-C. Based on the transformed $\tilde{\mathbf{x}}^{(m)}(l),\forall m,l$, in \eqref{eq05}, all the components from the odd rows of $\tilde{\mathbf{x}}^{(m)}(l)$ represent the symbol sent from one of relay nodes, and all the components from the even rows of $\tilde{\mathbf{x}}^{(m)}(l)$ represent the symbol sent from the source node. Let's define a modulated bits set $\mathcal{S}_{\mathrm{odd}}$ representing all the components from the odd rows of $\tilde{\mathbf{x}}^{(m)}(l)$, and $\overline{\mathcal{S}}_{\mathrm{odd}}$ is its complementary set representing all the components from the even rows of $\tilde{\mathbf{x}}^{(m)}(l)$. By taking $p^{(l)}_e$ into account, the \textit{a posteriori} probabilities of the modulated bits sent from the relay node can be modified as
\begin{IEEEeqnarray}{ll}
p(\hat{c}^{(l)}_{j,m}=+1)=(1-p^{(l)}_e)p(c^{(l)}_{j,m}=+1)+p^{(l)}_ep(c^{(l)}_{j,m}=-1),\nonumber\\
p(\hat{c}^{(l)}_{j,m}=-1)=(1-p^{(l)}_e)p(c^{(l)}_{j,m}=-1)+p^{(l)}_ep(c^{(l)}_{j,m}=+1),\nonumber
\end{IEEEeqnarray}
where $p(\hat{c}^{(l)}_{j,m})$ denotes the modified probability of the $j^{\mathrm{th}}$ bit sent from the relay node and $j\in\mathcal{S}_{\mathrm{odd}}$. Then, by replacing $p(c^{(l)}_{j,m})$ with $p(\hat{c}^{(l)}_{j,m})$ in \eqref{eq10} when $j\in\mathcal{S}_{\mathrm{odd}}$, the LLR value of the $i^{\mathrm{th}}$, $i\in\overline{\mathcal{S}}_{\mathrm{odd}}$, modulated bits sent from the source node can be expressed as
\begin{IEEEeqnarray}{ll}\label{eq11}
L[c^{(l)}_{i,m}|\tilde{\mathbf{y}}^{(m)}_{\mathrm{D}}(l)]=\mathrm{log}\frac{\sum_{\mathbf{x}:c^{(l)}_{i,m}=+1}e^{-\Delta^{(m)}(l)+\Theta^{(m)}(l)}}{\sum_{\mathbf{x}:c^{(l)}_{k,m}=-1}e^{-\Delta^{(m)}(l)+\Theta^{(m)}(l)}},
\end{IEEEeqnarray}
where
\begin{IEEEeqnarray}{ll}\label{eq12}
\Theta^{(m)}(l)\triangleq\sum_{j,j\in\overline{\mathcal{S}}_{\mathrm{odd}}}{\mathrm{log}p(c^{(l)}_{j,m})}+\sum_{j,j\in\mathcal{S}_{\mathrm{odd}}}{\mathrm{log}p(\hat{c}^{(l)}_{j,m})}.
\end{IEEEeqnarray}
On the other hand, due to the error prorogation effects, the LLR value of the $i^{\mathrm{th}}$, $i\in\mathcal{S}_{\mathrm{odd}}$, modulated bits sent from the relay node can be expressed as \eqref{eq13}, shown on the top of next page, 
\begin{figure*}
\normalsize \vspace*{-0pt}
\begin{IEEEeqnarray}{ll}\label{eq13}
L[c^{(l)}_{i,m}|\tilde{\mathbf{y}}^{(m)}_{\mathrm{D}}(l)]=\log\frac{p(c^{(l)}_{i,m}=+1)}{p(c^{(l)}_{i,m}=-1)}+\mathrm{log}\frac{(1-p^{(l)}_e)\sum_{\mathbf{x}:\hat{c}^{(l)}_{i,m}=+1}e^{-\Delta^{(m)}(l)+\tilde{\Theta}^{(m)}(l)}+p^{(l)}_e\sum_{\mathbf{x}:\hat{c}^{(l)}_{i,m}=-1}e^{-\Delta^{(m)}(l)+\tilde{\Theta}^{(m)}(l)}}{(1-p^{(l)}_e)\sum_{\mathbf{x}:\hat{c}^{(l)}_{i,m}=-1}e^{-\Delta^{(m)}(l)+\tilde{\Theta}^{(m)}(l)}+p^{(l)}_e\sum_{\mathbf{x}:\hat{c}^{(l)}_{i,m}=+1}e^{-\Delta^{(m)}(l)+\tilde{\Theta}^{(m)}(l)}},\nonumber\\
\end{IEEEeqnarray}
\vspace*{-0.2cm}\hrulefill 
\end{figure*}
where
\begin{IEEEeqnarray}{ll}\label{eq14}
\tilde{\Theta}^{(m)}(l)\triangleq\sum_{j,j\in\overline{\mathcal{S}}_{\mathrm{odd}}}{\mathrm{log}p(c^{(l)}_{j,m})}+\sum_{j,j\neq i\in\mathcal{S}_{\mathrm{odd}}}{\mathrm{log}p(\hat{c}^{(l)}_{j,m})}.
\end{IEEEeqnarray}
Here, we omit the detailed derivation of \eqref{eq13} because of the limited space. On the other hand, the basic principle of deriving \eqref{eq13} is to solve $\sum_{j,j\in\mathcal{S}_{\mathrm{odd}}}{\mathrm{log}p(\hat{c}^{(l)}_{j,m})}$ on the condition that $c^{(l)}_{i,m}=+1$ for the numerator of \eqref{eq13} and on the condition that $c^{(l)}_{i,m}=-1$ for the denominator of \eqref{eq13}. Similar analisis method for the network coding aided two-way relaying system can be found in \cite{Lee2009}.

\begin{figure}[t] 
\begin{center}
\epsfig{figure=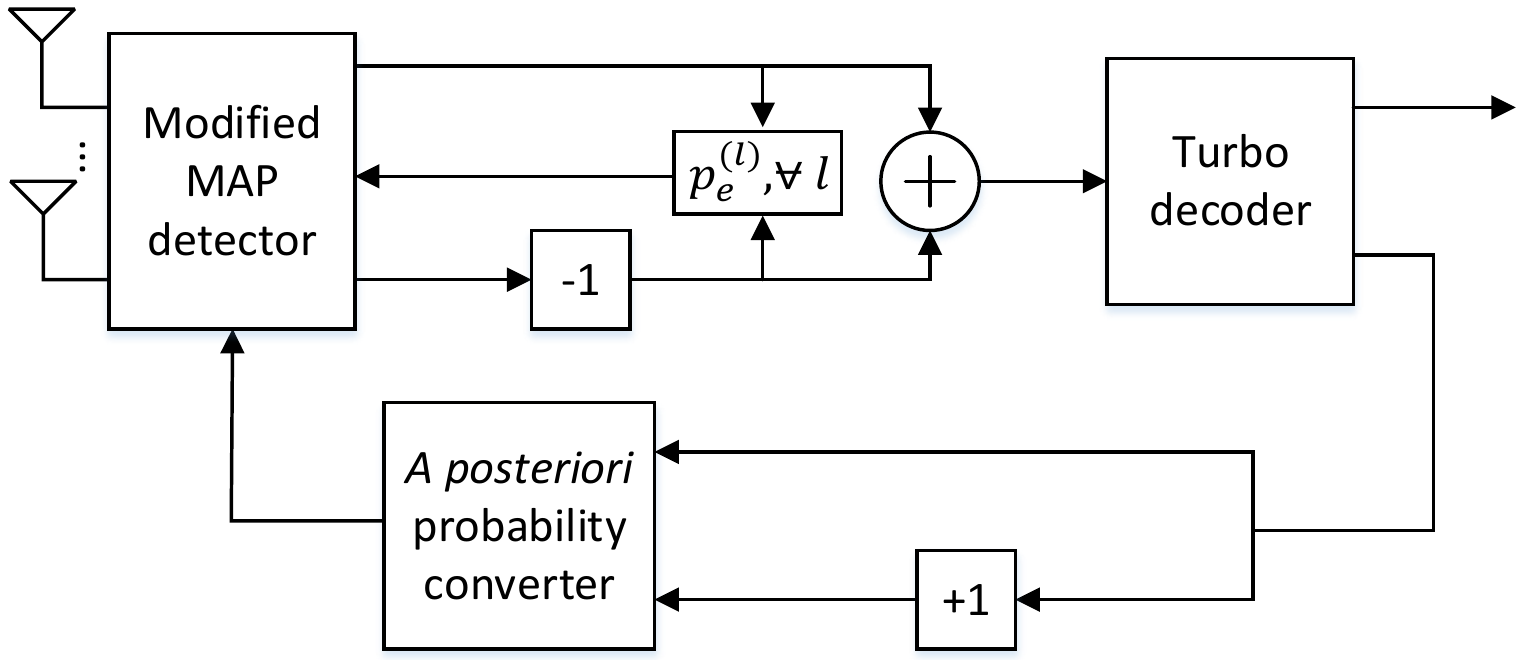,scale=0.55,angle=0}
\end{center}
\caption{Reliability-aware iterative decoding model.}\label{F2}
\end{figure}

Fig.~\ref{F2} illustrates the iterative detection/decoding process at destination node. 
With the obtained LLR values from \eqref{eq11} and \eqref{eq13}, the modified MAP detector now is able to separate the data frames sent from the relay and the source nodes, meanwhile, the error propagation effect can be eliminated by taking $p^{(l)}_{e},\forall l,$ into account. Here, destination node needs to wait all $L$ frames being detected by the modified MAP detector before the iterative process started. Then, it will add up two versions of LLR values per frame, e.g., the $(l-1)^{\mathrm{th}}$ frame sent from the relay node in the $l^{\mathrm{th}}$ time slot and the $(l-1)^{\mathrm{th}}$ frame sent from the source node in the $(l-1)^{\mathrm{th}}$ time slot, for the turbo-like decoder. The LLR values of the coded bits per frame, after the turbo decoder, will then be duplicated and fed into the \textit{a posteriori} probability converter to generate the \textit{a posteriori} probabilities for the next iteration. In the \textit{a posteriori} probability converter, the \textit{a posteriori} probabilities are found from the LLR values using the modulator mapping function. In addition, in Fig.~\ref{F2}, the blocks ``-1'' denote backward-shifting the frame number, ``+1'' denote forward-shifting the frame number, and $p^{(l)}_e,\forall l,$ denote the error probability parameter estimation.   
 
%%%%%%%%%%%%%%%%%%%%%%%%%%%%%%%%%%%%%%%%%%%%%%
\subsection{Error Probability Parameter Estimation}
As discussed in Section III-B, in order to mitigate the error propagation effect, the probabilities of decoding error at relay nodes $p^{(l)}_e,\forall l,$ should be taken into account in the iterative decoding process. These probabilities can be estimated at relay nodes and then be forwarded to destination node at the cost of signalling overhead. On the other hand, they can be directly estimated at destination node, where the estimation method will be introduced in this subsection.

As shown in Fig.~\ref{F2}, the error probability estimation process should be carried out in each of detection/decoding iterations. In detail, let's define $B\triangleq\log_{2}Q$ is the number of modulated bits per each constellation symbol sent from either source node or one of relay nodes. Then, the estimated probabilities can be given by
\begin{IEEEeqnarray}{ll}\label{eq15}
p^{(l)}_{e} = \frac{1}{MB}\sum^{M}_{m=1}\sum^{B}_{j=1}\mathrm{Pr}[c^{(l)}_{j,m}=+1|S]\mathrm{Pr}[c^{(l)}_{j,m}=-1|R]\nonumber\\
~~~~~~~~+\mathrm{Pr}[c^{(l)}_{j,m}=-1|S]\mathrm{Pr}[c^{(l)}_{j,m}=+1|R],~\forall l,
\end{IEEEeqnarray}
where $\mathrm{Pr}[c^{(l)}_{j,m}=\pm1|S]$ are the probabilities for the bits sent from the source node;  $\mathrm{Pr}[c^{(l)}_{j,m}=\pm1|R]$ are the probabilities for the bits sent from the relay node. Then, \eqref{eq15} can be further transformed to 
\begin{IEEEeqnarray}{ll}\label{eq16}
p^{(l)}_{e} = \frac{1}{MB}\sum^{M}_{m=1}\sum^{B}_{j=1}\frac{e^{L[c^{(l)}_{j,m}|S]}+e^{L[c^{(l)}_{j,m}|R]}}{(1+e^{L[c^{(l)}_{j,m}|S]})(1+e^{L[c^{(l)}_{j,m}|R]})},~\forall l,\nonumber\\
\end{IEEEeqnarray}
where $L[c^{(l)}_{j,m}|S]\triangleq\log\frac{\mathrm{Pr}[c^{(l)}_{j,m}=+1|S]}{\mathrm{Pr}[c^{(l)}_{j,m}=-1|S]}$ represents the LLR value of the modulated bit sent from the source node, which is derived from \eqref{eq11}; $L[c^{(l)}_{j,m}|R]\triangleq\log\frac{\mathrm{Pr}[c^{(l)}_{j,m}=+1|R]}{\mathrm{Pr}[c^{(l)}_{j,m}=-1|R]}$ represents the LLR value of the modulated bit sent from one of relay node, which is derived from \eqref{eq13}. After the error probabilities for all $L$ frames being estimated in each iteration, they will be fed back and used by the modified MAP detector in the next iteration. It is worth noting that the accuracy of the estimated probabilities can be improved with the iteration number.

%%%%%%%%%%%%%%%%%%%%%%%%%%%%%%%%%%%%%%%%%%%%%%
\subsection{Computational Complexity of Iterative Decoding}
Based on the above analysis, the proposed iterative decoding process can be summarized as
\begin{center}
{\normalsize
\begin{tabular}{cp{8cm}}
& \textit{Reliability-Aware Iterative Decoding}\\ \hline
&I.~ Initialize \textit{A posteriori} probabilities and set $i=0$;\\
&II.~\textbf{While} $i\leq Iter$:\\
&~~~~~$1.$ Increasing $i$ by 1;\\
&~~~~~$2.$ Calculate LLRs for all $L$ frames based on \eqref{eq11}\\
&~~~~~~~~ and \eqref{eq13};\\
&~~~~~$3.$ Estimate $p^{(l)}_e,\forall l,$ based on \eqref{eq16};\\ 
&~~~~~$4.$ Combine two version of LLR values per frame\\
&~~~~~~~~ and then send them to turbo decoder;\\
&~~~~~$5.$ Send the LLRs of coded bits to \textit{A posteriori}\\
&~~~~~~~~ probability converter;\\
&~~~~~$6.$ Feed back the output of \textit{A posteriori} probabili-\\
&~~~~~~~~ -ties to modified MAP detector;\\
&III.~ \textbf{End While}\\
\hline
\end{tabular}}
\end{center}
where $Iter=5$ is the number of outer iterations. The computational complexity mainly comes from Step II-2 and Step II-4. Specifically, for Step II-2, modified MAP detector is implemented and its complexity mainly depends on the number of receive antenna and modulation size; for Step II-4, the complexity of standard turbo-like decoder should be taken into account. For the rest steps, like Step II-3 and Step II-5, the complexity just involves certain element-wise multiplication and additions, which can be ignored by comparing with Step II-2 and Step II-4.  
%%%%%%%%%%%%%%%%%%%%%%%%%%%%%%%%%%%%%%%%%%%%%%%%%%%%%%%%%%%%%%%%%%%%%%
%%%%%%%%%%%%%%%%%%%%%%%%%%%%%%%%%%%%%%%%%%%%%%%%%%%%%%%%%%%%%%%%%%%%%%

                     %%%IV. Simulation Results%%% 

%%%%%%%%%%%%%%%%%%%%%%%%%%%%%%%%%%%%%%%%%%%%%%%%%%%%%%%%%%%%%%%%%%%%%%
\section{Simulation Results}
%Para. 1 introduces the simulation environment as well as your
%experimental design.
Computer simulations are used to evaluate the BER performances of our proposed multi-antenna assisted virtual FD relaying with reliability-aware iterative decoding method. We assume all channels are generated as independent block Rayleigh fading, which remain static over each transmission time slot.
For transmitting, source and relay nodes are using one transmit antennas, and for receiving, relay and destination nodes are with $N=2$ receive antennas. In addition, there are $L=20$ frames being transmitted via $L+1$ time slots, and each frame consists of $M=512$ information bits. SNR denotes the transmission signal power normalized by noise power ratio. The quadrature phase-shift keying (QPSK) modulation and turbo-like channel coding are implemented, e.g., 1/2 rate serial concatenated convolutional code for both source and relay nodes, where the first encoder is the non-recursive non-systematic convolutional code with a generator polynomial $G=([3,2])_8$, and the second encoder is the doped-accumulator with a doping rate equalling two. Random interleavers are implemented, and the results are computed on an average over 1000 independent channel realizations.

Four baselines are used for comparisons: 1) \textit{Perfect decoding at relay nodes}: this is served as performance bound, where we assume both relay nodes can correctly decode their received messages all the time and the same iterative decoding at destination node as our proposed scheme; 2) \textit{Proposed iterative decoding (ID) without $p_e$}: relay nodes always forward their decoded symbols regardless decoding errors and the iterative decoding at destination node does not take $p_e$ into account; 3) \textit{CRC based S-DF}: relay nodes only forward if their received messages can be correctly decoded and the same iterative decoding at destination node as our proposed scheme; 4) \textit{Threshold based S-DF}: relay nodes only forwards if their decoded errors are less than 15\% and the same iterative decoding at destination node as our proposed scheme. In addition, the QR-decomposition based detection method at relay nodes is implemented for all the schemes.

\begin{figure}[t] 
\begin{center}
\epsfig{figure=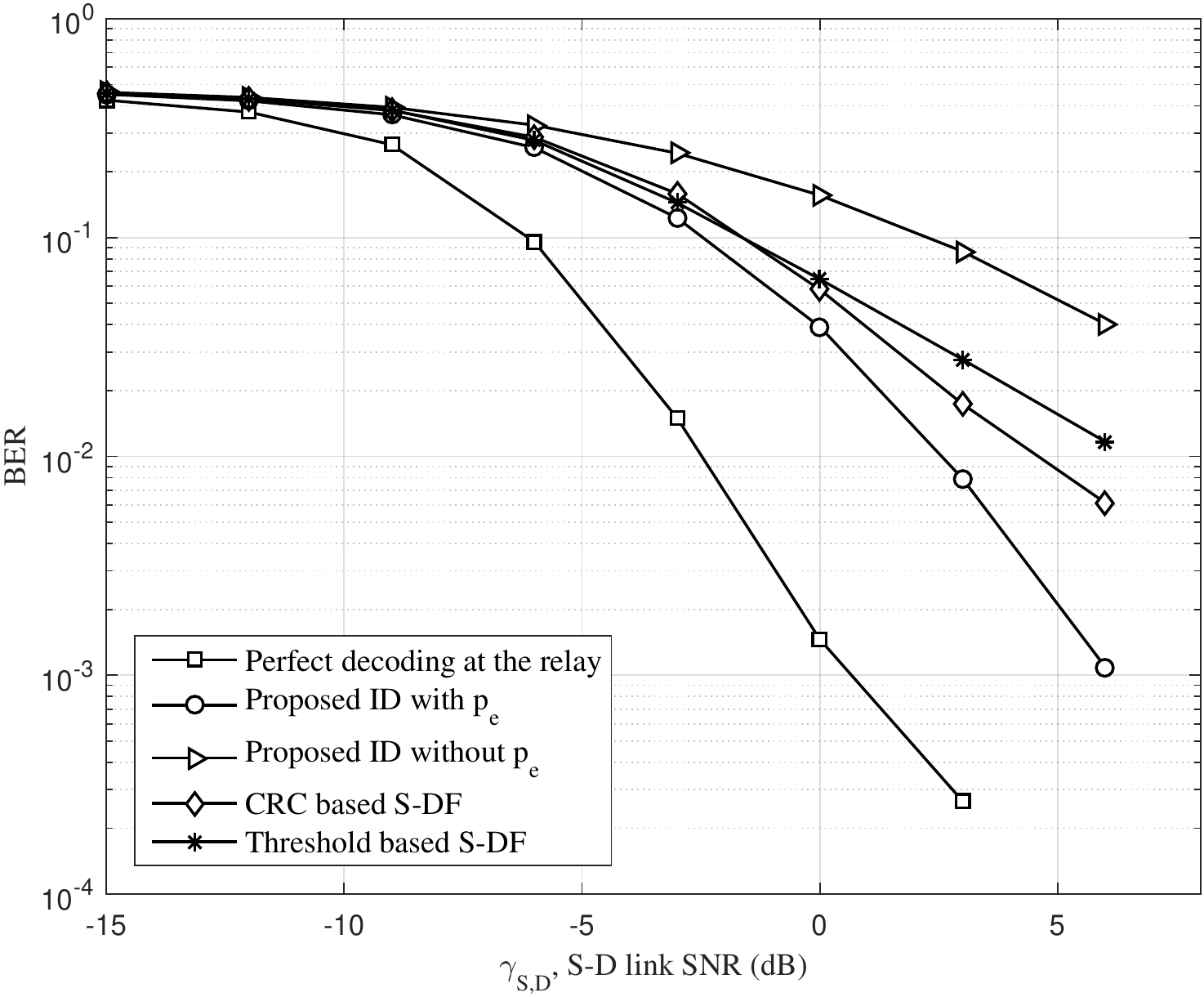,scale=0.53,angle=0}
\end{center}
\caption{BER versus S-D link SNR with the relay-location A in block Rayleigh fading channels and block-length=1024.}\label{F3}
\end{figure}

Fig.\ref{F3} shows BER performances for different schemes with relay-location A. 
Let's define $d_{i,j}$ as the distance between the $i^{\mathrm{th}}$ and $j^{\mathrm{th}}$ nodes, and $d_{\mathrm{S,D}}=d$. Thus, we have $d_{\mathrm{S,R1}}=d_{\mathrm{S,R2}}=d_{\mathrm{R1,D}}=d_{\mathrm{R2,D}}=d$ for the relay-location A. As a consequence, based on a simplified suburban area pathloss model with the pathloss exponent equalling to 3.52, SNR relationship in dB among different links can be approximated by $\gamma_{\mathrm{S,D}}=\gamma_{\mathrm{S,R1}}=\gamma_{\mathrm{S,R2}}=\gamma_{\mathrm{R1,D}}=\gamma_{\mathrm{R2,D}}$. As shown in Fig.\ref{F3}, \textit{Perfect decoding at relay nodes} scheme undoubtedly gives the best BER performance. \textit{Proposed ID with $p_e$} scheme gives the second best BER performance due to the iterative decoding process at destination node by taking the error propagation effects into account. In contrast, \textit{Proposed ID without $p_e$} scheme gives the worst BER performance. The BER performances of two S-DF based schemes are in the middle between \textit{Proposed ID with $p_e$} scheme and \textit{Proposed ID without $p_e$} scheme. This is because the S-DF based schemes prefer to remain silent rather than propagate decoding errors if the number of decoding errors above their limits.   

\begin{figure}[t] 
\begin{center}
\epsfig{figure=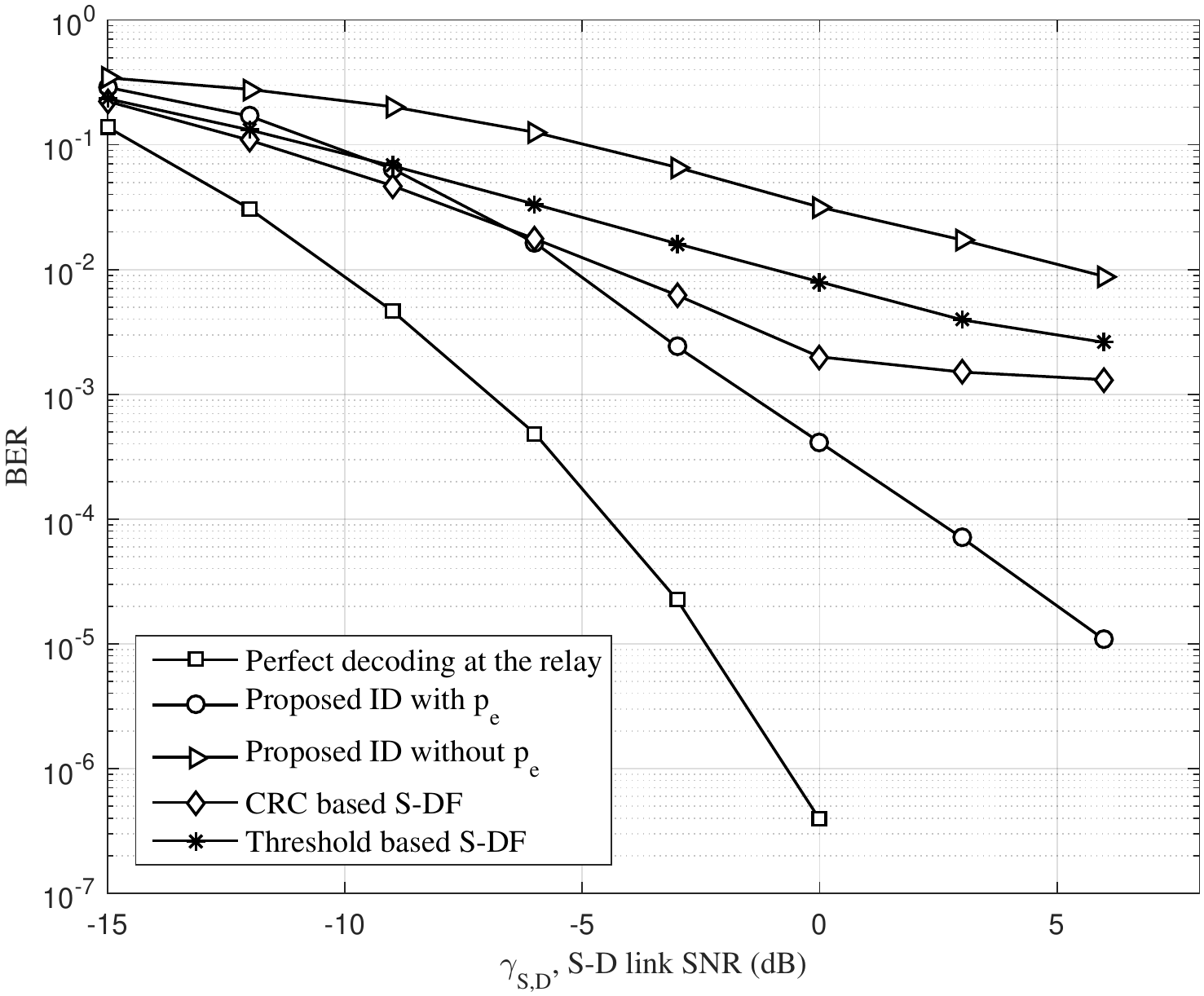,scale=0.53,angle=0}
\end{center}
\caption{BER versus S-D link SNR with relay-location B in block Rayleigh fading channels and block-length=1024.}\label{F4}
\end{figure}

Fig.\ref{F4} gives BER performances for different schemes with a different relay location (e.g. relay-location B).
In this case, given $d_{\mathrm{S,D}}=d$, we have $d_{\mathrm{S,R1}}=d_{\mathrm{S,R2}}=\frac{1}{2}d$ and $d_{\mathrm{R1,D}}=d_{\mathrm{R2,D}}=\frac{3}{4}d$. Then, similar as relay-location A, the corresponding SNR relationship in dB among different links can be approximated by $\gamma_{\mathrm{S,R1}}=\gamma_{\mathrm{S,R2}}=\gamma_{\mathrm{S,D}}+10.6~\mathrm{dB}$ and $\gamma_{\mathrm{R1,D}}=\gamma_{\mathrm{R2,D}}=\gamma_{\mathrm{S,D}}+4.4~\mathrm{dB}$. Such location is the case that relay nodes are close to source node. As shown in Fig.\ref{F4}, the same BER performance trend as Fig.\ref{F3} is presented, where \textit{Proposed ID with $p_e$} scheme gives the second best BER performance. In comparison with Fig.\ref{F3}, the BER performance gap between \textit{Proposed ID with $p_e$} scheme and the conventional S-DF based schemes in Fig.\ref{F4} is larger than the ones in Fig.\ref{F3}, where the performance gap in Fig.\ref{F4} exhibits up to 8 dB gain. This is because, with relay-location B, the decoding capability of relay nodes increases due to high S-R link quality. On the other hand, for the conventional S-DF based schemes, one or a few bits decoding errors will stop them forwarding their received frames to enjoy the spatial diversity gain. This is also why there is error floor for \textit{CRC based S-DF} scheme.

%Other paras. detail your experimental report.
%%%%%%%%%%%%%%%%%%%%%%%%%%%%%%%%%%%%%%%%%%%%%%%%%%%%%%%%%%%%%%%%%%%%%%

                         %%% VI. Conclusion %%%

%%%%%%%%%%%%%%%%%%%%%%%%%%%%%%%%%%%%%%%%%%%%%%%%%%%%%%%%%%%%%%%%%%%%%%
\section{Conclusion}
In this paper, a multi-antenna assisted virtual FD relaying with reliability-aware iterative decoding has been proposed. Based on QR decomposition, our proposed scheme can cancel the inter-relay interference without boosting noise power at relay nodes. In addition, our proposed scheme allows two relay nodes forward the erroneously decoded symbols and jointly cancels inter-frame interference and error propagation effect at destination node. It has been shown that, without extra cost of time delay and signalling overhead, our proposed scheme exhibits up to 8 dB gain by comparing with the conventional S-DF based schemes especially when relay nodes are close to source node.

%%%%%%%%%%%%%%%%%%%%%%%%%%%%%%%%%%%%%%%%%%%%%%%%%%%%%%%%%%%%%%%%%%%%%%

                         %%%Acknowledgment%%%

%%%%%%%%%%%%%%%%%%%%%%%%%%%%%%%%%%%%%%%%%%%%%%%%%%%%%%%%%%%%%%%%%%%%%%
\section*{Acknowledgement}
%The authors would like to thank...% the Editor Prof. and anonymous reviewers for their efficient review process and constructive comments.
This work was supported by the Engineering and Physical Science Research Council (EPSRC) through the Scalable Full Duplex Dense Wireless Networks (SENSE) grant EP/P003486/1.

%%%%%%%%%%%%%%%%%%%%%%%%%%%%%%%%%%%%%%%%%%%%%%%%%%%%%%%%%%%%%%%%%%%%%%
%\section*{Appendix I}
%Put the proof of main result here.

\bibliography{main}
\bibliographystyle{IEEEtran}
\end{document}